\documentclass[traditabstract]{aa}
\usepackage[pdftex]{color,xcolor}
\usepackage[pdftex]{graphicx}
\usepackage{natbib}
\bibpunct{(}{)}{;}{a}{}{,}
\usepackage{float}
\usepackage{stfloats}
\usepackage{fixltx2e}
\usepackage{txfonts}
\usepackage{wasysym}
\begin{document}
   \title{The HARPS search for southern extrasolar planets. }
          \subtitle{XXIX. Four new planets in orbit around the moderatly active dwarfs    
                  \object{HD\,63765},  \object{HD\,104067},
                  \object{HD\,125595}, and \object{HIP\,70849}.
                  \thanks{The {\footnotesize HARPS} radial velocity measurements discussed in this paper are only available in electronic form
at the CDS via anonymous ftp to cdsarc.u-strasbg.fr (130.79.128.5) or via http://cdsweb.u-strasbg.fr/cgi-bin/qcat?J/A+A/}
                  \thanks{Based on observations made with the {\footnotesize HARPS}
                          instrument on the ESO 3.6 m telescope at La Silla
                          Observatory under the GTO programme ID 072.C-0488.}
                  }   
\author{D. S\'egransan\inst{1}
        \and M.~Mayor\inst{1} 
        \and S.~Udry\inst{1}
        \and C.~Lovis\inst{1}
        \and W-~Benz\inst{2} 
        \and F.~Bouchy\inst{3,4}  
        \and G.~Lo Curto\inst{7}   
        \and C.~Mordasini\inst{2,5}
        \and C.~Moutou\inst{6}    
        \and D.~Naef\inst{1} 
        \and F.~Pepe\inst{1} 
        \and D.~Queloz\inst{1} 
        \and N.~Santos\inst{8}  
      } 
   \offprints{Damien S\'egransan, \email{Damien.Segransan@unige.ch}}
   \institute{Observatoire Astronomique de l'Universit\'e de Gen\`eve, 
                   51 ch. des Maillettes - Sauverny -, CH-1290 Versoix, 
                   Switzerland
          \and
          Physikalisches Institut, Universitat Bern, Silderstrasse 5, CH-3012 Bern, Switzerland
          \and
          Institut d'Astrophysique de Paris, UMR 7095 CNRS, Universit\'e Pierre \& Marie Curie, 98bis boulevard Arago, 75014 Paris, France 
          \and
          Observatoire de Haute-Provence, 04870 Saint-Michel l'Observatoire, France
          \and
          Max-Planck-Institut fŸr Astronomie, K\"onigstuhl 17, D-69117 Heidelberg  , Germany  
          \and
          Laboratoire d'Astrophysique de Marseille, UMR 6110 CNRS, UniversitŽ de Provence, 38 rue Fr\'ed\'eric Joliot-Curie, 13388 Marseille Cedex 13, France
          \and
          European Southern Observatory, Karl-Schwarzschild-Str. 2, D-85748 Garching bei M\"unchen   , Germany            
          \and
          Centro de Astrof\'{\i}sica , Universidade do Porto, Rua das Estrelas, 4150-762 Porto, Portugal 
	     }

   \date{Accepted}

  \abstract   { We report the detection of four new extrasolar planets in orbit around the
    moderately active stars \object{HD\,63765}, \object{HD~104067}, \object{HIP~70849}, and \object{HD~125595} 
    with the {\footnotesize HARPS} Echelle spectrograph 
    mounted on the ESO 3.6-m  telescope at La Silla.
    The first planet, \object{HD~63765}~b,  has a minimum mass of
    0.64\,M$_{\rm Jup}$,  a period of 358 days, and an eccentricity of
    0.24. It orbits a G9 dwarf at 0.94 AU.  
    The second planet, \object{HD~104067}~b, is a
    3.6 Neptune-mass-planet with a 55.8-day-period. It orbits its parent K2 dwarf, in a circular orbit with a
    semi-major axis of   $a$=0.26\,AU. Radial velocity measurements present a $\approx$500-day-oscillation that reveals 
     significant magnetic cycles. 
    The third planet is a 0.77 Neptune-mass-planet in circular orbit around the K4 dwarf, \object{HD~12595}, with a 9.67-day-period.
    Finally, \object{HIP~7849\,b} is a  long-period (5$<P<$75\,years) and massive planet of $m.\sin{i} \approx$3.5-15\,M$_{\rm Jup}$ that orbits a late K7 dwarf.
   }
   

   \maketitle
%________________________________________________________________

\section{Introduction}
   The {\footnotesize HARPS} planet-search  program 
   has been going on for 6 years (since 2003) 
   at the 3.6-m ESO telescope located at La Silla Observatory,
   Chile \citep{Pepe-2000,Mayor-2003:a}. 
   Both \object{HD\,63765} and \object{HD\,104067}
   were part of the high-precision subprogram of the  {\footnotesize HARPS} Guaranted
   Time Observions (GTO),
   which aimed at detecting very low-mass planets in a sample 
   of solar-type stars already screened for giant planets at a lower 
   precision with  {\footnotesize CORALIE} Echelle spectrograph
   mounted on the 1.2-m Swiss telescope on the 
   same site  \citep{Udry-2000:a}.  
   \object{HIP\,70849} and \object{HD\,125595}, on the other hand, are part of the HARPS volume-limited sample program of the  
    {\footnotesize HARPS} GTO. The sample is composed of about 1200 stars  within 57.5pc, among which 850 G, K  dwarfs that are known as single stars
     with low activity and low $v.\sin{i}$.  The goal of this program was to monitor the 850 stars  at the 3 ms$^{-1}$ precision level to 
     detect all the gas giants present in this statistically robust sample.  Observed distributions of planets' orbital elements and masses, 
      as well as host star characteristics should then be compared to theoretical predictions to improve our knowledge of planet formation and evolution.

\begin{figure*}[t!]
\center
        \includegraphics[angle=0,width=0.80\textwidth,origin=br]{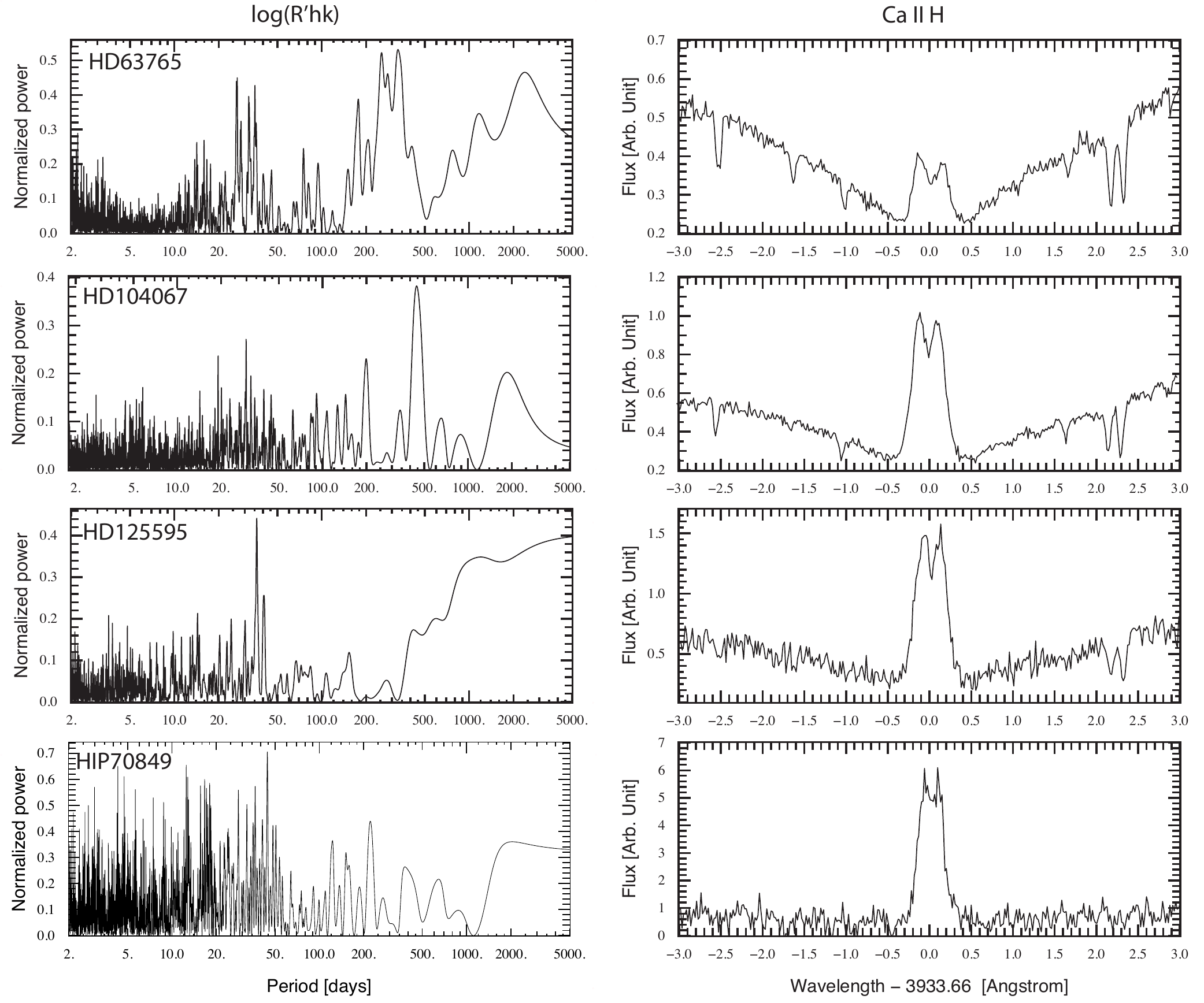}
  \caption[]{
\label{fig:periodigram_rhk}
\label{fig:CaII}
Left : Chromospheric activity index  periodogram for {\footnotesize HD}\,63765, {\footnotesize HD}\,104067, 
{\footnotesize HD}\,125595, and {\footnotesize HIP}\,70849 . Right : Ca II H emission region. The large re-emission at the bottom of the Ca II H 
absorption line  at $\lambda=3933.66~\AA$ is an indicator of chromospheric activity. }
\end{figure*}

So far,  {\footnotesize HARPS}  has allowed the detection (or has
contributed to the detection) of more than 85 extrasolar planet candidates
\cite[the latest batch of which was presented in a series of parallel papers e.g.][]{Forveille-2011, Lovis-2011,  Mordasini-2011, Moutou-2011, LoCurto-2010, Naef-2010} among the $\approx$460  known today.
 But more specifically, {\footnotesize HARPS}
is unveiling the tip of the iceberg of the very low-mass planet
population \cite[see ][]{Lovis-2009a, Mayor-2009} which
when combined with forthcoming results allow extrapolating results of their
 mass distribution. \citet{Lovis-2009a} show that $\approx$30\% of the 
stars surveyed in the {\footnotesize HARPS} high precision program 
have planets with masses below 30-earth masses on periods shorter than 50
days. Furthemore, $\approx$80\% of those planets are in multplanetary systems.
 {\footnotesize HARPS}' high stability not only gives access to 
 super-Earth's at short periods but is also sensitive to
 Neptune-mass-planets at a longer period, \cite[e.g., HD~69830~b,
 ][]{Lovis-2006}. The discovery of those intermediate-mass planets
 (1-10 M$_{\rm Nept}$) in the  0.1 to 0.5\,AU separation range is of interest for 
planetary formation scenarios since those planets are hardly
formed according to \citet{Ida-2005}
or at least, should be rare according to \citet{Mordasini-2009b}.

 In this paper we report the detection of four planets orbiting moderately active stars. Two of them are Neptunian planets at relatively short periods orbiting  
 \object{HD\,104067} and  \object{HD\,125595}. The other  two are gas giants at longer periods 
 in orbit around   \object{HD\,63765} and  \object{HIP\,70849}.
The paper is organized as follows. In the first section 
we briefly discuss the primary-star properties. Data analysis is presented in section \ref{sec:genetic} while 
radial-velocity measurements and  orbital solutions are discussed  
in section \ref{sec:measurements}. In section \ref{sec:discussion}, 
we provide  concluding remarks.

\section{Stellar characteristics}
\label{sec:characteristics}  
The ffective temperatures, gravity, and metallicities were derived from 
 the spectroscopic  analysis of the {\footnotesize HARPS}  targets as described in  \citet{Sousa-2008}. 
 We used the improved  Hipparcos astrometric  parallaxes rederived by \citet{vanLeeuwen-2007} to determine the 
absolute V-band magnitude using the apparent visual magnitude from Hipparcos \citep{ESA-1997}.
Metallicities, together with the effective temperatures and $M_{V}$ were then used to estimate basic 
stellar parameters (ages, masses) bases on theoretical isochrones from \cite{Girardi-2000} 
and a Bayesian estimation method described in \cite{daSilva-2006}\footnote{The web interface for the Bayesian estimation of stellar parameters, called  {\footnotesize PARAM 1.0}, can be found at  http://stev.oapd.inaf.it/cgi-bin/param. }.
Individual spectra were also used to derive both
the bisector inverse slope ( {\footnotesize BIS} ) of the   {\footnotesize HARPS} 
cross-correlation function ( {\footnotesize CCF} ), as defined by \citet{Queloz-2000a0}, 
as well as a measurement of the chromospheric activity index $\log{\left(R^{'}_{HK}\right)}$, 
following a similar recipe as used by \citet{Santos-2000:b} for  {\footnotesize CORALIE} spectra.
Using the\citet{Noyes-84} empirical calibration of the  rotational period
vs. activity,  we estimated  rotational periods.  We also derived  the $v.\sin{(i)}$ from the  
{\footnotesize HARPS}  spectra  following the \citet{Santos-2002} approach.
The HIP\,70849 spectral  type is, however, too late (K7V) to carry out  \citet{Sousa-2008}'s analysis. For such a low-mass star,%
we have to rely on near-infrared photometric measurements, as well as on accurate mass-luminosity relations \citep{Baraffe-1998, Delfosse-2000}, to derive its fundamental physical parameters. The list of all stellar parameters can be found in Table~\ref{tab:StellarParam}.

The four stars have activity indexes, $\log{\left(R^{'}_{HK}\right)}$, ranging between -4.69 and -4.80,
and they show a large Ca II re-emission at  $\lambda=3933.66~\AA$ (see Fig. \ref{fig:CaII}), revealing significant chromospheric activity possibly induced by stellar spots or plagues. 
In addition, $\log{\left(R^{'}_{HK}\right)}$ indexes display
variability on different time-scales as observed on their  periodograms (see Fig. \ref{fig:periodigram_rhk}).
 Short time-scale variation can often
 be matched to the rotation period of the star (such as for  \object{ {\footnotesize HD}\,63765} and \object{ {\footnotesize HD}\,125595} ),
  while longer time scale variability in the chromospheric activity index could betray the existence of stellar magnetic cycles
  (such as for  \object{ {\footnotesize HD}\,63765}, \object{ {\footnotesize HD}\,104067} and \object{ {\footnotesize HD}\,125595} ). 
   \object{{\footnotesize HIP}\,70849}  does not display any significant variabilitiy in the $\log{\left(R^{'}_{HK}\right)}$ index.

For some stars, a linear correlation between the activity index and the radial velocities is observed (see Fig. \ref{fig:rv_hd104067}), 
leading the way to possible stellar activity detrending around moderately active stars. 
However, this topic is beyond the scope of this paper so we decided, instead, to 
 estimate the overall stellar radial velocity jitter. To do so, we measured the radial velocity dispersion on a sample of
stars with similar $\log{\left(R^{'}_{HK}\right)}$ and spectral type that show no planetary. % signature with more than 10 measurements..
For each star, a radial velocity jitter of a few m.s$^{-1}$ was derived (see Table~\ref{tab:StellarParam}), which was used as 
an external white noise in the planet search algorithm (see Sect. \ref{sec:genetic}).

\begin{table*}[t!]
\caption{\label{tab:StellarParam}
Observed and inferred stellar parameters for \object{HD\,63765},\object{HD\,104067}, \object{HIP\,70849} and \object{HD\,125595}.}
\begin{center}
\begin{tabular}{llcccccc}
\hline\hline
Parameters                         &             &\object{HD~63765}         &\object{HD~104067}   &\object{HD~125595}  &\object{HIP~70849}       \\[1mm]
\hline                                                        
Sp. T.                                    &             &G9V                                         &K2V                                        &   K4V       &K7V        \\ [1mm]
V                                            &             &8.1                                            &7.93                                             &   9.03  &10.36\\[1mm]
$B-V$                                   &             &0.75                                          &0.99                                               &   1.107 &  1.427	  \\[1mm]
J                                            &             &      -                                             &              -                                               &  -    &7.639$\pm$0.023 \\[1mm]
H                                            &             &    -                                               &             -                                                &   -  &7.006$\pm$0.061\\[1mm]
K                                           &             &        -                                            &             -                                          &  -&6.790$\pm$0.027  \\[1mm]
$\pi$                                     & [mas]       &30.07 $\pm$	0.56               &47.47 $\pm$0.90   & 35.77	$\pm$ 1.04 &42.42$\pm$2.09     \\[2mm] 
\hline
 $M_{V}$                            &              &5.49$\pm$0.04                     &6.31$\pm$0.05             &6.80$\pm$0.06 &8.50$\pm$0.10         \\[1mm]
 $T_{\rm eff}$                         & [K]          &5432$\pm$19$^{(1)}$       &4969$\pm$72$^{(1)}$            &4908$\pm$ 	87$^{(1)}$ &4105 $\pm$130$^{(4)}$    \\[1mm]
$[Fe/H]^{(1)}$                          & [dex]             &-0.16$\pm$0.01    &-0.06$\pm$0.05      &$0.02\pm$0.06&-   -\\[1mm]
$\log{(g)}^{(1)}$&&4.42$\pm$0.03&4.47$\pm$0.13& 4.32 $\pm$0.20&-\\
$M_{\star}$                        &[M$_{\odot}$]  &0.865$\pm$0.029$^{(3)}$           & 0.791$\pm$0.020$^{(3)}$      & 0.756 $\pm$ 0.017$^{(3)}$&0.63$\pm$0.03$^{(4)}$             \\[1mm]
Age                                     &[Gyr]  &4.77$\pm$4.07$^{(3)}$        &4.33$\pm$3.99$^{(3)}$            &   3.53$\pm$ 3.59 $^{(3)}$&1-5$^{(4)}$ \\[1mm]
$\log{\left(R^{'}_{HK}\right)}$   &              & -4.736$\pm$0.037$^{(2)}$                &  -4.743$\pm$0.023$^{(2)}$        &  -4.787$\pm$0.037$^{(2)}$ &-4.697$^{(5)}$\\[1mm]
Stellar Jitter              &    [m/s]    &   3.0 &   3.0       &  2.5  &3.5\\[1mm]

$v\sin{(i)}^{(2)}$&[km\,s$^{-1}$]&1.63&1.61& 1.50&1.93\\
$P_{rot}$&[days]&$~$23&$~$34.7&40.1&-\\[1mm]
\hline
\end{tabular}
\tablefoot{ Near infrared photometry is taken from the 2MASS All-Sky Catalog, \citep{Cutri-2003,Skrutskie-2006}. Astrometric data and visual photometry come from from Hipparcos Catalogs,  \citep{ESA-1997,vanLeeuwen-2007}. Other parameters come from:(1) \citet{Sousa-2008} spectroscopic analysis, (2) { \footnotesize HARPS CCF}, (3) \citet{Girardi-2000} models, (4) \citet{Baraffe-1998} models, (5) \citet{Gray-2006}.}
\end{center}
\end{table*}

\section{Radial-velocity measurements  and orbital solutions}
\label{sec:measurements}

\subsection{A Bayesian genetic algorithm with Markof chains}
\label{sec:genetic}
Since 2003, we have been developing and using a genetic algorithm to retrieve multiple Keplerian orbital solutions 
 \citep[see][in prep]{Segransan-2011:a}. The main advantage of genetic algorithm compared to other advanced methods such as "Markov chain Monte Carlo" (hereafter MCMC) with parallel tempering \citep[for illustration see][]{Gregory-2007} is that it allows probing the full model parameters space in a extremely efficient way (solutions of complex planetary systems solutions are derived in a few minutes instead of hours to run MCMC from scratch on a MacBook Pro). However, the population at the end of the evolution is not statistically reliable owing the intrinsic nature of genetic algorithm based of genome crossover and mutation. We therefore added a MCMC module with Metropolis hasting to overcome this problem. The algorithm is based on the formalism described in \citet{CollierCameron-2007} with several thousands chains drawn from the final genetic algorithm's population. Each chain runs several thousand times to retrieve a statistically reliable population. For poorly constrained planetary systems, the Bayesian formalism described in \citet{Gregory-2007} is used. A full overview of the algorithm will be described in a forthcoming paper \citep[see][in prep]{Segransan-2011:a}.

 \subsection{A one-year period Jovian planet around  {\footnotesize HD}\,63765 }
\begin{figure}[h!]
        \includegraphics[angle=0,width=0.40\textwidth,origin=br]{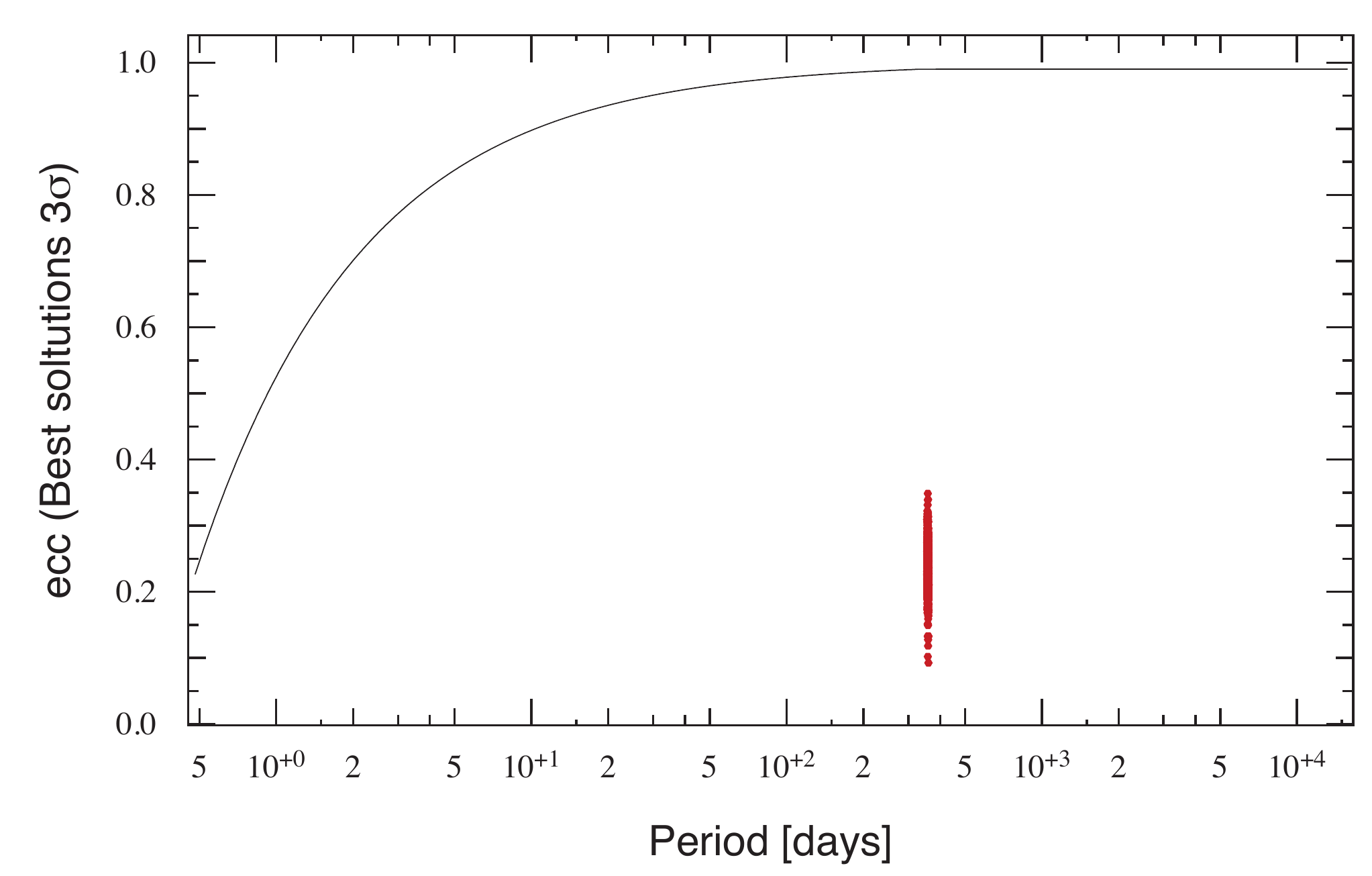}
        \includegraphics[angle=0,width=0.40\textwidth,origin=br]{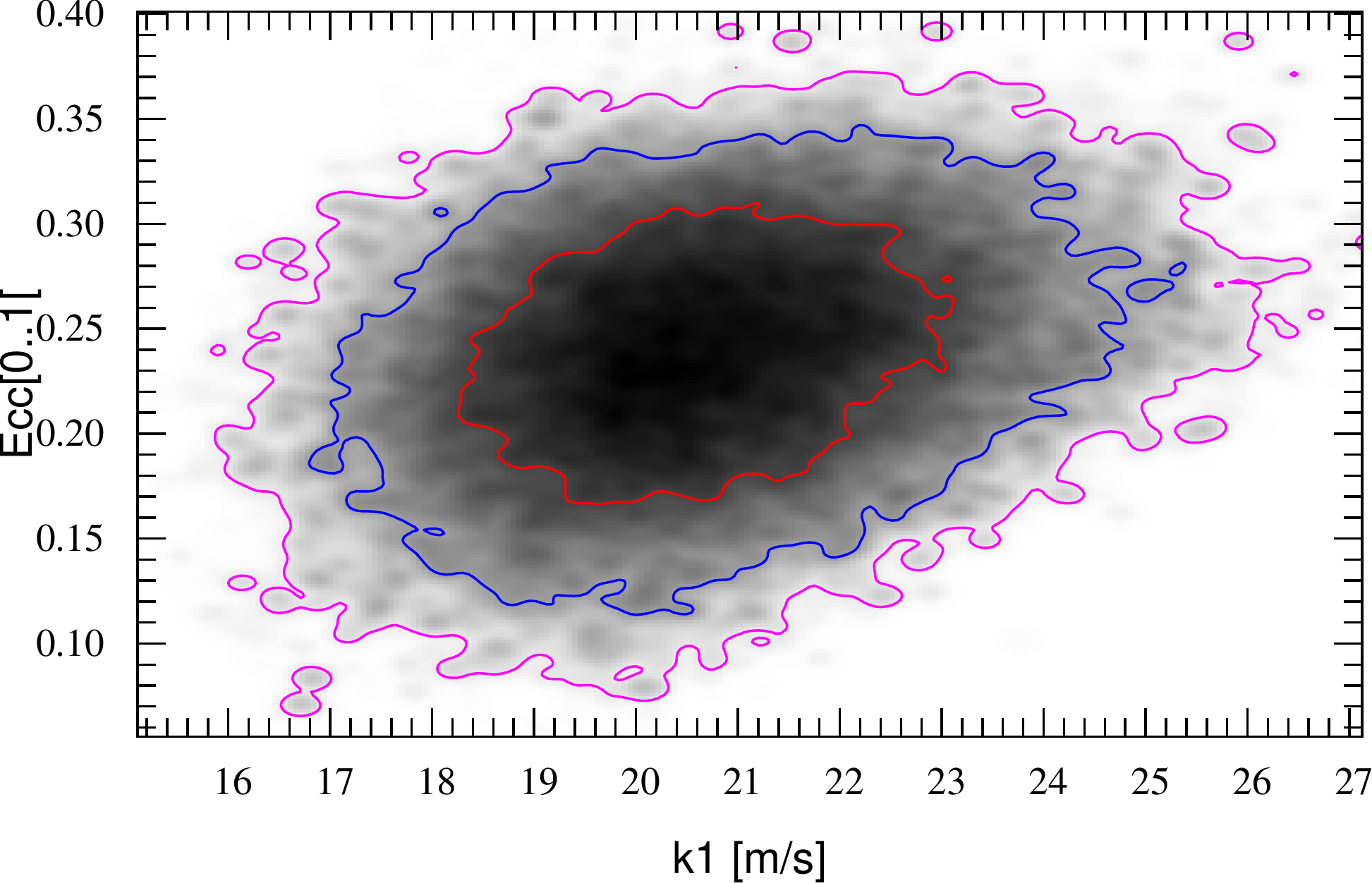}
  \caption[]{
\label{fig:stat_hd63765}
Top:genetic algorithm population at the end of evolution represented in a period-eccentricity diagram. The dark curve is the maximum eccentricity a planet can reach without colliding the star. Bottom:joint probability density function of the radial velocity semi-amplitude and the eccentricity. The red, blue and cyan contour represent the 1, 2 and 3 $\sigma$ confidence intervals of the joint PDF.}
\end{figure}

\begin{figure*}[t!]
\center
 \includegraphics[angle=0,width=0.80\textwidth,origin=br]{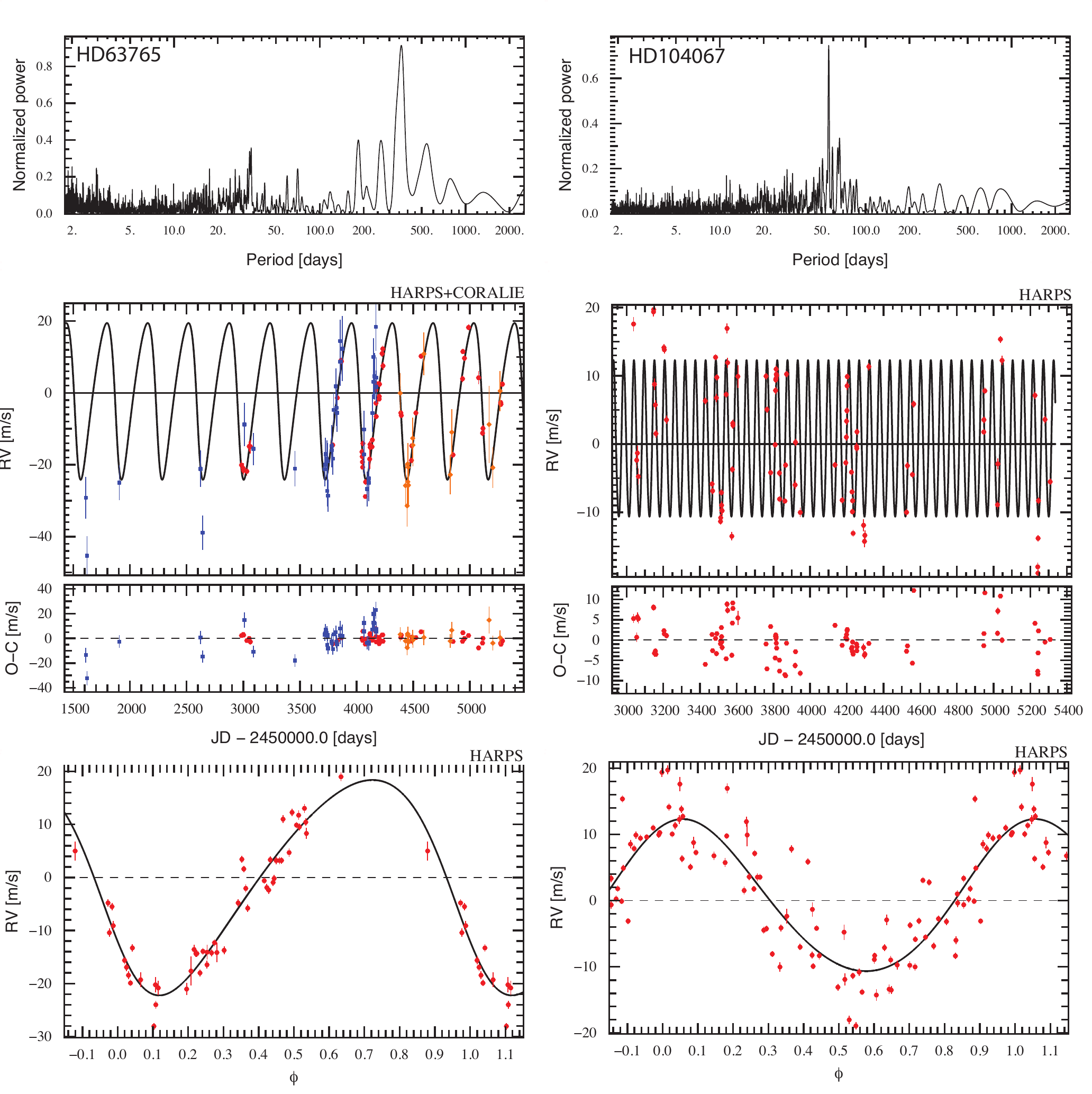}\\
  \caption[]{
\label{fig:rv_hd63765}
\label{fig:rv_hd104067}
Top:Generalized Lomb-Scargle periodogram of the radial-velocity measurements for both HD\,63765 (left) and HD\,104067 (right). 
Middle:Radial-velocity measurements as a function of Julian Date obtained with {\footnotesize  {\footnotesize CORALIE} }(blue and orange dots) and
{\footnotesize  {\footnotesize HARPS} }(red dots). 
 {\footnotesize  {\footnotesize HARPS'} } only phase-folded radial velocities are also displayed (bottom).
 The best Keplerian, one-planet-solution is displayed as a dark curve whose corresponding orbital elements are listed in Table~\ref{tab:InstPerf}.
}
\end{figure*}

 \object{HD~63765} was part of the {\footnotesize CORALIE} planet
 search survey \citep{Udry-2000:a} before being monitored at 
higher precision with  {\footnotesize HARPS}.
Since March 2000, 56 measurements have been acquired with
 {\footnotesize CORALIE}  with a typical signal-to-noise ratio of 46 
(per pixel at 550 nm) and a mean measurement uncertainty (including photon noise and calibration errors) of
4.5\,ms$^{-1}$. In June 2007, we
upgraded the instrument that slightly changes its zero point \citep{Segransan-2009:a}. We
therefore adjusted an instrumental offset between the measurements taken 
before and after June 2007 (namely  {\footnotesize CORALIE-98} and {\footnotesize CORALIE-07}).
Since December 2003, we have recorded  a total of 52 measurements
with {\footnotesize HARPS} with a better signal-to-noise (SNR$\approx$155) than
{\footnotesize CORALIE} as well as a better final accuracy
(60\,cm$^{-1}$, 
including photon noise and calibration errors). 
Despite the lower accuracy of the {\footnotesize CORALIE} spectrograph, we decided to include the {\footnotesize CORALIE} data because of their
 longer time span, mainly to increase the accuracy on the
orbital period of the planet.
Altogether, 108 radial velocities  were 
gathered. A stellar velocity jitter of 3.0\,m.s$^{-1}$ was quadratically added to the mean radial velocity uncertainty. 
A single-planet Keplerian model was adjusted to the data  and produced residuals that show a level
   of variation $\sigma$=3.41\,m.s$^{-1}$, yielding a reduced $\chi^{2}$ of 1.30.
Figure \ref{fig:rv_hd63765} shows the  {\footnotesize HARPS}  
radial velocities  and the  corresponding best-fit Keplerian model.
 The resulting orbital parameters for the first planet are $P$=358\,days, $e$=0.24 and $K$=20.9\,ms$^{-1}$ , 
implying a minimum mass $m.\sin{(i)}$=0.64\,M$_{\rm Jup}$
  and a semi-major axis $a$=0.940\,AU.
  Despite its proximity to one year, we are confident that the 358-day period is the correct one. Indeed, the genetic algorithm does not display any other significant solution in a period-eccentricity diagram (see Fig. \ref{fig:stat_hd63765}). This is reasonably explained by the large radial-velocity amplitude ($K =20.9\,$ms$^{-1}$) compared to the residual noise ($\sigma$=3.41\,m.s$^{-1}$) and the fact that 75\% of the orbital phase is covered (see Fig. \ref{fig:rv_hd63765}). In addition, the mcmc analysis shows that the probability distribution functions (hereafter PDF) of all parameters are gaussian with no obvious correlations between them as illustrated by the semi-amplitude vs. eccentricity diagram (see Fig. \ref{fig:stat_hd63765}). Orbital elements for  \object{HD~63765}~b are listed in Table \ref{tab:OrbitalParam}.

\begin{table}[th!]
\caption{
\label{tab:InstPerf}
Number of measurement, time span, and weighted r.m.s. of the residuals
around the one planet solutions for \object{HD~63765} for {\tiny CORALIE-98}, {\tiny CORALIE-07},  and {\tiny HARPS}.  }
\begin{center}
\begin{tabular}{lccc}
\hline\hline
Instrument          &          $N_{mes}$              &  {\it Span} &  $\sigma_{(O-C)}$   \\
         &                       & [years]  & [ms$^{-1}$]    \\
\hline                                                                                         \\[-2mm]                         
    {\tiny CORALIE-98}            & 38    &    7.0   &  9.6   \\    [1mm]
   {\tiny CORALIE-07}            & 18    &    3.18   &  4.72   \\    [1mm]
   {\tiny HARPS}            & 52    &    6.3   & 3.27   \\    [1mm]
\hline 
\hline
\end{tabular}
\end{center}
\end{table}

\subsection{A 55-day-period Neptune-like planet around  {\footnotesize HD}\,104067 }

\begin{figure}[h!]
        \includegraphics[angle=0,width=0.40\textwidth,origin=br]{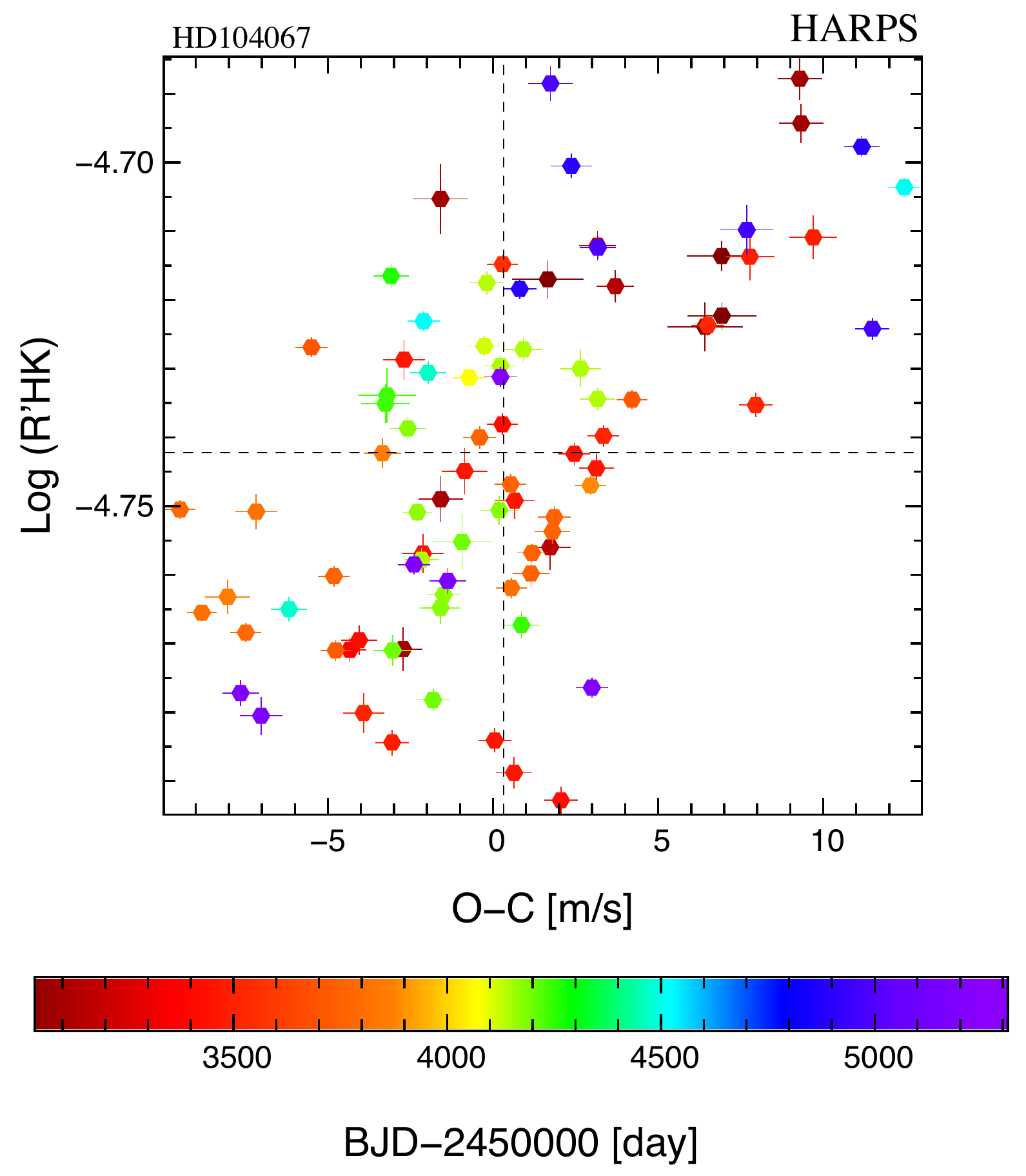}
  \caption[]{
\label{fig:Correl}
Correlation bewtween the activity index and the radial velocity residuals as function of time for {\footnotesize HD}\,104067. Pearson's r = 0.59.
}
\end{figure}

 \object{HD~104067}  has been observed with  {\footnotesize HARPS} since February 2004. Altogether, 88 radial-velocity 
measurements with a typical signal-to-noise ratio of $\approx 172$ 
(per pixel at 550 nm) and a mean measurement uncertainty (including photon noise and calibration errors) of 43\,cms$^{-1}$ were 
gathered.  A stellar velocity jitter of 3.0\,m.s$^{-1}$ was quadratically added 
to the mean radial velocity uncertainty. 

A strong peak is present in the periodogram at $\approx$56\,day (see Fig. \ref{fig:rv_hd104067}), which is identified as a 
Neptune-like planet in a circular orbit by the genetic algorithm. We investigated whether   the presence of stellar spots could mimic the short amplitude
  $P = 55.81$\,day Keplerian motion  and explain such a large
  $\chi^{2}_{r}=2.46$ since  the derived orbital period  - from activity index - is 34.7 days.  
  The bissector inverse slope, as well as the $\log{\left(R^{'}_{HK}\right)}$ (see Fig. \ref{fig:periodigram_rhk}),  
  do not present any variation at the period of P=55.81\,day,  
  while the measured radial velocities show a clear Keplerian signature. The detection of the  $55.81$-day-period Neptunian planet is therefore confirmed.
  
The residuals' dispersion is fairly large ($\sigma=4.6$\,m.s$^{-1}$)
and shows a clear low frequency oscillation at P$\approx$500\,day that mimics the signature of a second Neptunian planet.
However, this signal is correlated with the $\log{\left(R^{'}_{HK}\right)}$ activity index (see Fig. \ref{fig:Correl}), ruling out the  detection of a second long period planet in the system. 
Resulting orbital parameters are listed in Table \ref{tab:OrbitalParam}.
 
\subsection{A 10-day-period Netpunian planet around  {\footnotesize HD}\,125595 }
 We have been observing \object{HD~125595}  with  {\footnotesize HARPS} since May 2004. 
 Altogether, we have obtained 117 radial-velocity 
measurements with a typical signal-to-noise ratio of $\approx 78$
(per pixel at 550 nm) and a mean measurement uncertainty 
(including photon noise and calibration errors) of 1\,ms$^{-1}$. 

\begin{figure}[h!]
      \includegraphics[angle=0,width=0.45\textwidth,origin=br]{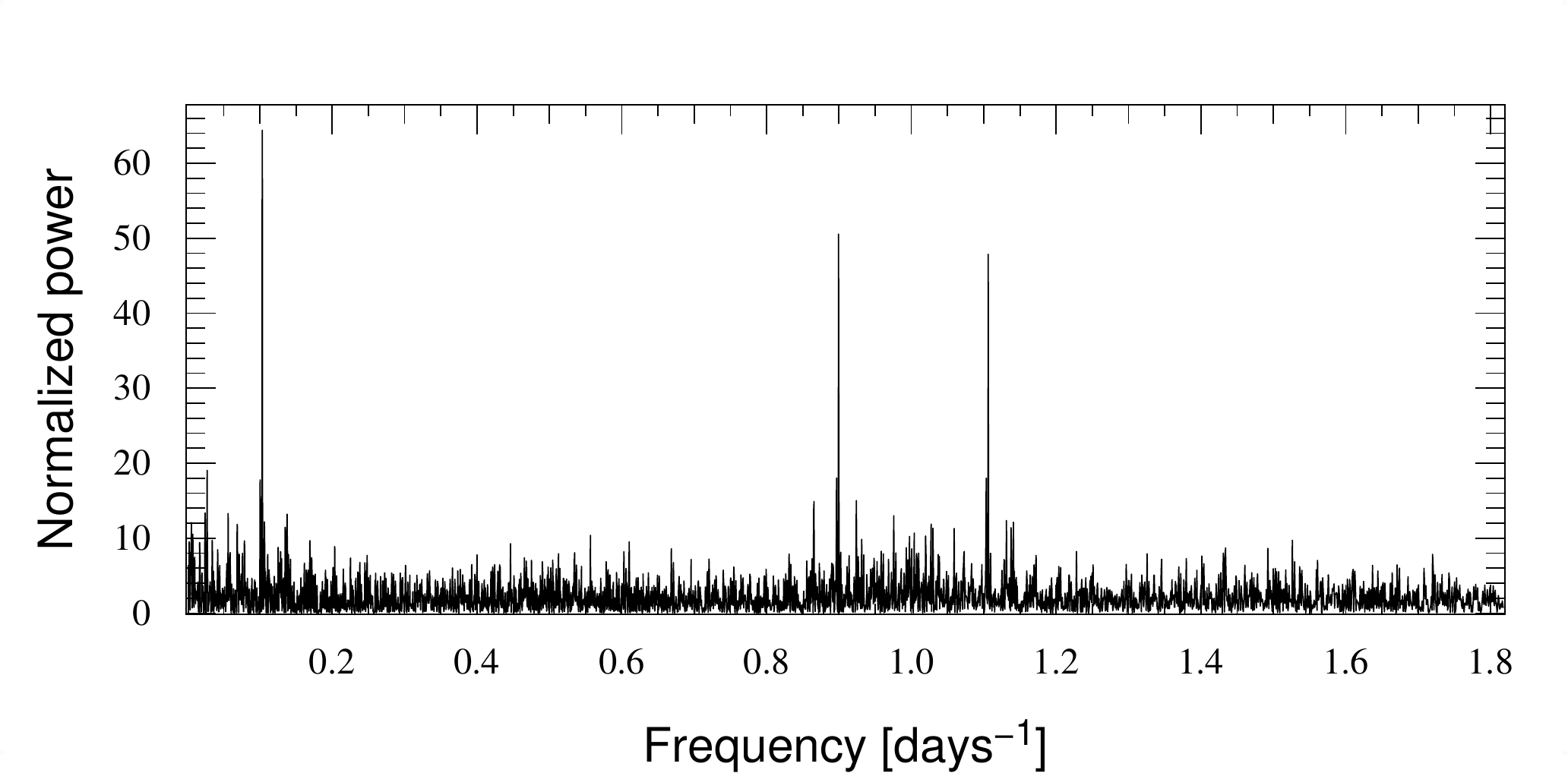} 
       \includegraphics[angle=0,width=0.45\textwidth,origin=br]{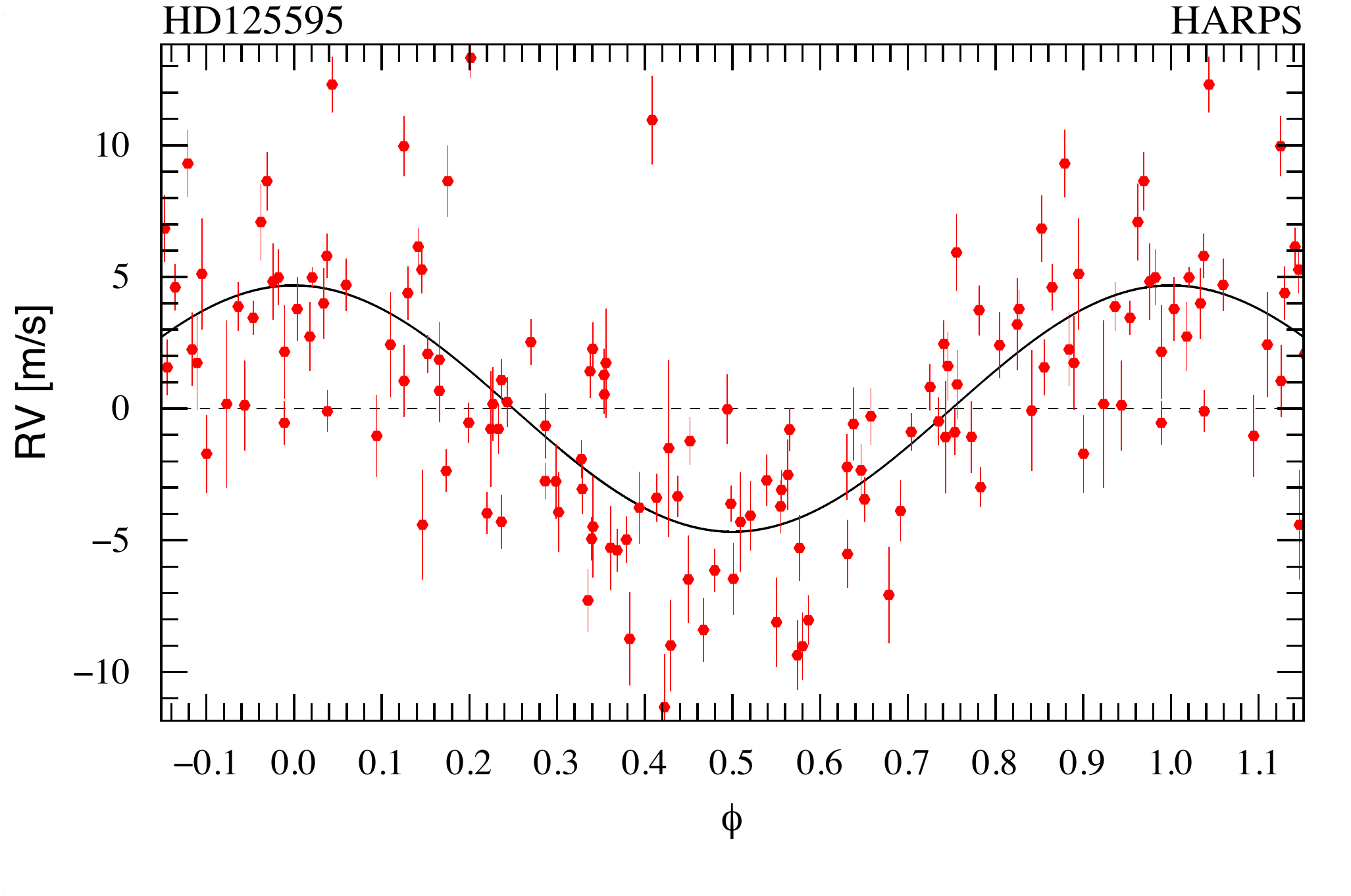} 

        \caption[]{
\label{fig:rv_hd125595}
 The top figure is the periodogram of {\footnotesize  {\footnotesize HARPS} } radial velocity measurements of HD125595. A clear signature is seen at frequency $\nu$=0.9 day$^{-1}$  with its replica on both sides of the one-day alias.
The bottom  figure represents the phase-folded  radial velocity measurements with the
 best Keplerian, one-planet-solution represented as a black curve.
}
\end{figure}

\begin{figure}[h!]
      \includegraphics[angle=0,width=0.45\textwidth,origin=br]{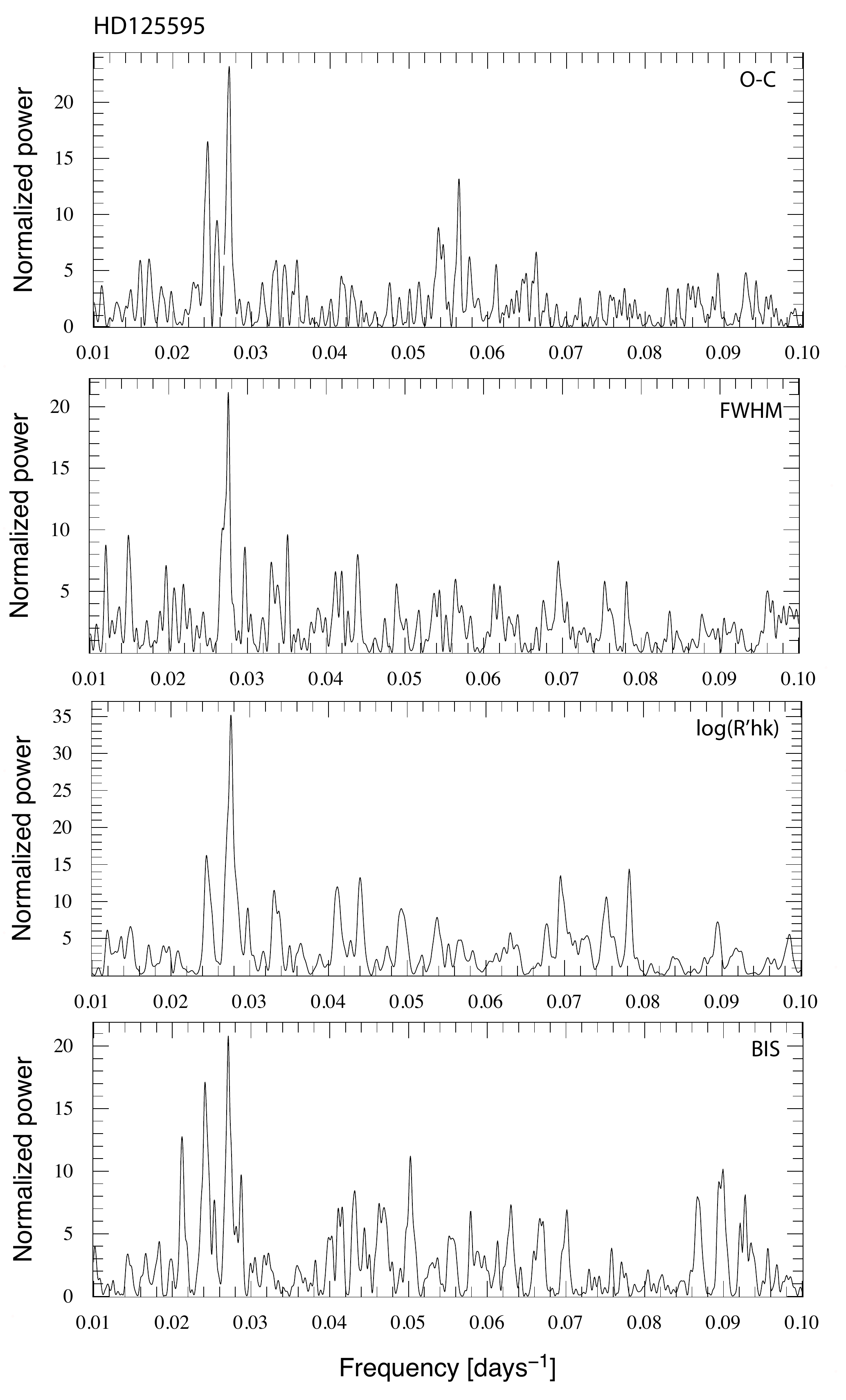} 
   \caption[]{
\label{fig:act_hd125595}
  The top figure is the periodogram of the radial velocities residuals (once the 9.67-day-period planet has been subtracted) on the time interval 10-100\,days.
   Some residual energy remains close to frequency (resp. period) $\nu$=0.027\,day$^{-1}$ (resp. P=37\,dayday). This energy peak is close to the 40-day-rotation period derived from the activity index,
   and is also present in the periodograms of the {\footnotesize CCF FHWM}  (resp.$\log{\left(R^{'}_{HK}\right)} $ and bisector inverse slope ) to state that the observed 37-day variability is activity induced.
}
\end{figure}

A clear $\approx$ 10-day periodic signal  can be seen in the periodogram ($\nu$=0.1 day$^{-1}$) with its replica on both sides of the one-day period alias (see Fig. \ref{fig:rv_hd125595}).
A single-planet Keplerian model was therefore adjusted to the data leading to a
$P=9.67$\,days circular orbit of semi-amplitude $K=4.79$\,ms$^{-1}$  (see Fig. \ref{fig:rv_hd125595}), which 
corresponds to a planetary companion of minimum mass $m\,\sin{(i)}=0.77$\,M$_{\rm Nept}$. Corresponding orbital elements are listed in Table \ref{tab:OrbitalParam}.
The $\sim3\,$ms$^{-1}$ dispersion in the residuals is explained by a significant amount of energy in the periodogram of the residuals around the orbital solution at a period of 37 days  (see Fig. \ref{fig:act_hd125595}). This $P\sim37$-day energy peak is also present in the periodograms of the CCF FHWM (resp. $\log{\left(R^{'}_{HK}\right)} $ and bisector inverse slope and is close enough to the rotation period of the star to asses that the observed dispersion in the radial velocities is probably induced by activity.

 \subsection{A long-period massive planet around  {\footnotesize HIP}\,70849 }

\begin{figure}[h!]
   \includegraphics[angle=0,width=0.45\textwidth,origin=br]{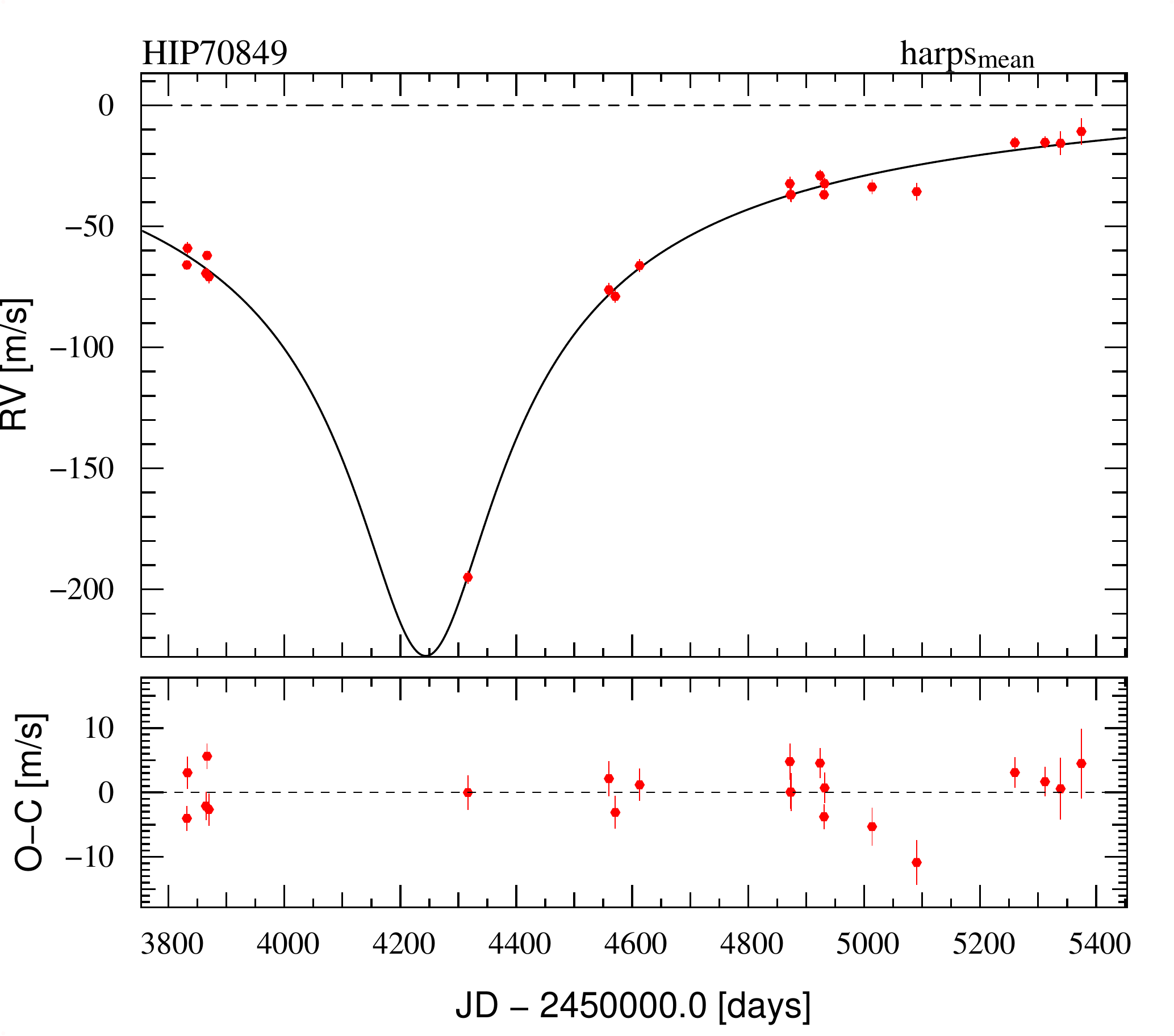}
\caption[]{
\label{fig:rv_hip70849}
Radial-velocity measurements  as a function of Julian Date obtained
with {\footnotesize  {\footnotesize HARPS} }(red dots) on {\footnotesize
  HIP}\,70849. One of the orbital solution is overplotted as a black line.
}
\end{figure}
 
\begin{figure}[h!]
   \includegraphics[angle=0,width=0.45\textwidth,origin=br]{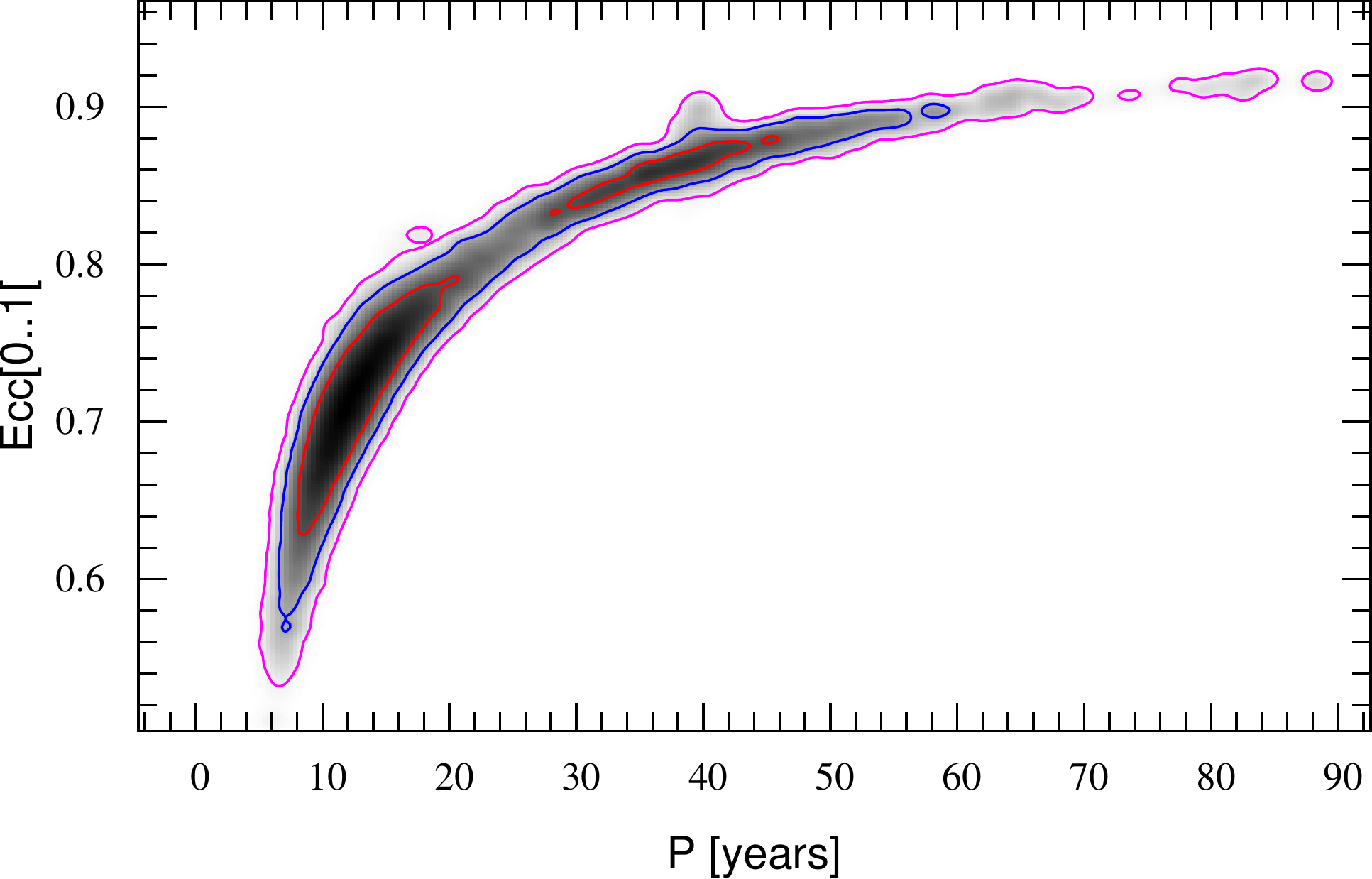}
   \includegraphics[angle=0,width=0.45\textwidth,origin=br]{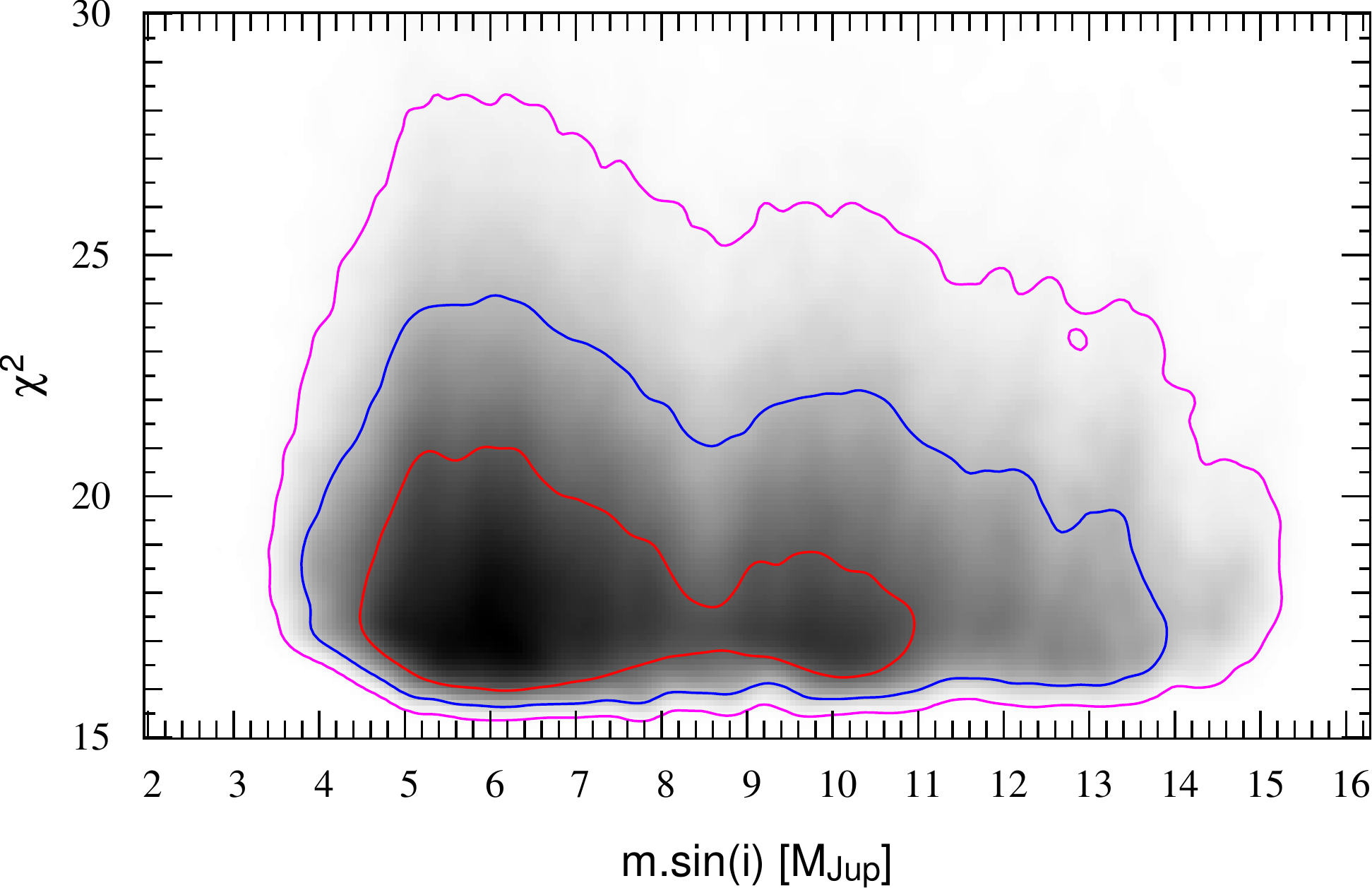} 
\caption[]{
\label{fig:gen_hip70849}
The top diagram represents the joint probability density function of the orbital period and the eccentricity. The red (resp. blue, cyan) line corresponds to the 1(resp. 2, 3)\,$\sigma$ confidence interval. The bottom diagram represents the minimum mass distribution of the planetary companions with the same confidence intervals. 99.73\% of the solutions have a period between 5 and 90\,years, an eccentricity between 0.47 and 0.95 and a minimum mass between 3.5 and 15\,M$_{\rm Jup}$. 
}
\end{figure}

 \object{HIP~70849}  has been observed with  {\footnotesize HARPS} since April 2006. Altogether, we gathered 18 radial-velocity 
measurements with a typical signal-to-noise ratio of $\approx 49$
(per pixel at 550 nm) and a mean measurement uncertainty 
(including photon noise and calibration errors) of 2.2\,ms$^{-1}$. 
A low frequency radial velocity variation with a large amplitude ($\approx 200\,$ms$^{-1}$) is clearly seen in the data (see Fig. \ref{fig:rv_hip70849}).
Even though the orbital phase is not correctly covered by our observations, we were lucky enough to obtain radial velocity 
measurements close to the date of passage through periastron. Running our genetic algorithm followed by MCMC simulations with the simplest model 
(1 planet, no radial velocity drift) allowed us to bring some constraints on the orbit as shown in Fig. \ref{fig:gen_hip70849}.
Indeed, at the end of the simulation the chains remained within bounding values with a period 
between 5 and 90 years, an eccentricity between 0.4 and 0.98 and a minimum mass between 3.5 and 15\,M$_{\rm Jup}$. It should be noted that solutions with much 
longer periods and very high eccentricities exist but could not be probed by our MCMC simulations in a reasonable amount of time. However, the corresponding minimum mass 
of such solutions remains within the substellar domain.
The addition of a linear drift to the model decreases the planet eccentricity, as well as its radial velocity semi-amplitude, which as a result, decreases the mass of the detected object. 
We also looked at the Hipparcos data to check whether  the astrometric signature of the companion ($\alpha>500\,\mu$arcsec) is detected, to determine the real nature of this object as for  \object{HD 5388}\,b \citep{Sahlmann-2011b}. In principle, astrometric motions larger that 1\,mas could be detected in the Hipparcos intermediate data as shown by \citet{Sahlmann-2011a}, provided that the RV orbital elements are determined well  and that the period is close to the duration of the Hipparcos mission. In that particular case, the long period of the planet combined with the poor determination of the orbital elements prevents to disentangle the companion signature from  Hipparcos proper motions. However, future space missions such as GAIA \citep{Perryman-2001} or JMAPS \citep{Hennessy-2010} will be able to characterize the nature of this object better.

 \begin{table*}[th!]
\caption{
\label{tab:OrbitalParam}
{\footnotesize  Best Keplerian orbital solutions for  \object{HD~63765},\object{HD~104067},  \object{HD~105595}, and    \object{HIP~70849}. }
}
\begin{center}
\begin{tabular}{llcccc}
\hline\hline
Parameters          &                       &      \object{HD~63765\, b}   &      \object{HD~104067\, b}    & \object{HD~125595\, b}            &              \object{HIP70849\, b }                 \\
\hline                                                                                         \\[-2mm]                         
$\gamma$            &  [kms$^{-1}$]                 &  $ 22.5843 \pm0.0024$& $15.1243\pm0.0009$ &5.2637$\pm$0.00054&     0.099$\pm $0.032 \\  [1mm]
$\Delta V({\tiny C07}-{\tiny C98})$&  [ms$^{-1}$]    &$-3.9\pm 2.4$    &&     \\  [1mm]
$\Delta V({\tiny H}-{\tiny C98})$&  [ms$^{-1}$]       &$39.4\pm $2.0    &&     \\  [1mm]
\hline                                                                                         \\[-2mm]                         
$t_{\rm{avg}}$         &{\tiny [BJD-2.45\,$10^{6}$]} &4224.7731&4001.5984&4439.9245&4564.01764\\
\hline                                                                                             
$P$                 &  [days]                    &   $358.0 \pm 1.0$ &  $55.806\pm 0.049$         & 9.6737  $\pm$  0.0039 &5-90 [years]\\  [1mm]
$K$                 & [ms$^{-1}$]                &   $20.9 \pm 1.3$    &    $11.56\pm 0.75$   & 4.79 $\pm$ 0.47                                                                                                                &70-400 \\  [1mm]
$e$                 &                              &  $0.240\pm0.043$  &     0.00     & 0.00                                                                                                                         & 0.47-0.96\\  [1mm]
 $\omega$           & [deg]                          &   $122\pm13$ &      0.00      &0.00                                                                                                                               & -\\  [1mm]
$T_{0}$             &{\tiny  [BJD-2.45\,$10^{6}$]   }   &    $4404\pm11$       &  $4043.15\pm0.56$    &4435.23$\pm$0.15                                                                     & -\\[1mm]
$\lambda(t_{\rm{avg}})$           &[deg]           &    $ 302.1\pm3.6$  &     $91.9\pm3.6$            &175$\pm$5   & -\\[1mm]                                                                  $\Lambda(t_{\rm{avg}})$           &[deg]           &    $302.3\pm7.0$  &     $91.9\pm3.6$            &175$\pm$5                                                                   & -\\[1mm]

 \hline                                                                  
$a_{1}\, \sin{i}$   & [$10^{-6}\, $AU    ]         &  $667.8\pm40$  &       $59  \pm3.7$                  &4.2 $\pm$ 0.41                                                                                               &$ 11.7-661\,.\,10^{3}$ \\ [1mm]
$f(m)$              & [$10^{-12}\, $M$_{\odot}$  ]     & $313\pm57 $  &       $9.0  \pm1.7$            &0.111  $\pm$0.031                                                                                       &   $ 95-428\,.\,10^{3}$\\ [1mm]
 $m\,\sin{i}$  &                & $0.64\pm0.05$\,M$_{\rm Jup}$   &     $3.44 \pm0.25$\,M$_{\rm Nept}$ &0.766 $\pm$0.078 \,M$_{\rm Nept}$  &  3-15\,M$_{\rm Jup}$  \\ [1mm]
  $a$               & [AU    ]         & $0.940\pm 0.016 $     &         $0.2643 \pm0.0045$         & 0.0809  $\pm$ 0.0014                      &4.5-36 \\      [1mm]                                                 
\hline                    
$N_{\rm{mes}}$              &             &  108      &   88           &117                                                                                                                                                                                                                    
&          21 \\ [1mm]
Time span         & [years]          &  10.77      &  6.22               & 5.68                                                                                                                                                                                       &        4.22\\ [1mm]
${\sigma_{(O-C)}}$   & [ms$^{-1}$]         &3.41    &   4.60            & 3.22                                                                                                                                                                                    & 3.94 - 3.68   \\    [1mm]      
$\chi^{2}_{r}$        &         &  $1.30\pm 0.16$    &    $ 2.46\pm 0.24$     & 1.61 $\pm$ 0.17  & 1.06 $\pm$ 0.38     \\              [1mm]                                                        
\hline       
\hline
\end{tabular}
\tablefoot{$t_{\rm{avg}}$ is the mean date of the observations, $\lambda(t_{\rm{avg}})$ is the mean longitude at $t=t_{\rm{avg}}$, 
             $\Lambda(t_{\rm{avg}})$ is the true longitude at  $t=t_{\rm{avg}}$, 
            ${\sigma_{(O-C)}}$ is the dispersion of the residuals.}
\end{center}
\end{table*}

\section{Conclusion}
\label{sec:discussion}   
 We have reported the discovery of four extrasolar planet candidates orbiting moderately active stars 
 -   \object{HD\,63765}, \object{HIP\,70849}, \object{HD\,104067}, and \object{HD\,125595} -
 discovered with the  {\footnotesize HARPS}  Echelle spectrograph mounted on the 3.6-m ESO telescope 
 located at La Silla Observatory.
Two of those planets,   \object{HD\,125595\,b}  and \object{HD\,104067\,b}, are Neptune-mass-planets while the other two - \object{HD\,63765}, \object{HIP\,70849} - are clearly gas giants.\\
 \\
 -- {\it \object{HD\,125595\,b}}:According to \citet{Mordasini-2009b}, the planet is too massive ($m_{p}=0.77\,M_{\rm Nept}$) to have formed ``in situ'' since there are not enough planetesimals at 0.08\,AU with typical disk profiles. Because its parent star has solar metallicity, the planet probably formed beyond the iceline and then migrated toward the star through tidal interaction with the gaseous disk. The planet is not massive enough to trigger runaway gas accretion. \\
--  {\it \object{HD\,104067\,b}}:The existence of such a planet, however, is hardly explained by planetary formation and evolution scenarios \citep{Ida-2005,Mordasini-2009b}. Indeed, the planet is massive enough ($m_{p}$=3.4\,M$_{\rm Nept}$) to have triggered runaway gas accretion before migrating toward the star. How could the detection of such a planet be explained? Are we biased by the observing techniques since \object{HD\,104067\,b} could be a gas giant provided that its inclination is lower that 30 $\deg$?  This question will be answered when such an object is detected in transiting surveys or when a statistically reliable sample of similar object is detected by radial velocity surveys.\\
-- {\it \object{HD\,63765\,b}}:This is a one-year-period Saturn-mass planet with a significant eccentricity. There is nothing peculiar about this object but its detection is useful to build a statistically robust sample of giant planets properties (orbital elements and mass distributions, host star properties).\\
-- {\it \object{HIP\,70849\,b}}:Interesting is the presence of a massive giant planet around the 0.63 solar-mass-star \object{HIP\,70849}. Even though characterized as a K7 dwarf,  \object{HIP\,70849} is close enough to the very low-mass star domain to present similar planetary formation scenarios. According to \citet{Laughlin-2004}, \citet{Ida-2005}, and \citet{Mordasini-2009b}, the formation of Jupiter-mass planets orbiting very low-mass stars is seriously inhibited at all separations.
The presence of this 3-15\,M$_{\rm Jup}$ planet in orbit to \object{HIP\,70849} brings to four the number of companions more massive than 2\,M$_{\rm Jup}$ in orbit around a very low mass star and therefore questions the validity of planets formation scenarios around very low-mass stars. Possible hints of the existence of such objects could be that the mass of their protoplanetary disk did not scale with the mass of the star or, otherwise, that the planet started to form much earlier than usually thought.\\

\begin{acknowledgements}
 We are grateful to the staff of the Geneva
Observatory, in particular to L.Weber, for maintaining the 1.2-m Euler
Swiss telescope and the CORALIE echelle spectrograph at La Silla,
and for technical support during observations.
 We thank the Swiss National Research
Foundation (FNRS) and the Geneva University for their continuous
support of our planet search programs.
NCS would like to acknowledge the support by the European Research Council/European Community under the FP7 through a Starting Grant, as well as from Funda\c{c}\~ao para a Ci\^encia e a Tecnologia (FCT), Portugal, through program Ci\^encia\,2007, and in the form of grants PTDC/CTE-AST/098528/2008 and PTDC/CTE-AST/098604/2008.  This research  made use of the SIMBAD database and of the VizieR catalog access tool operated at the CDS, France. This publication makes use of data products from the Two Micron All Sky Survey, which is a joint project of the University of Massachusetts and the Infrared Processing and Analysis Center/California Institute of Technology, funded by the National Aeronautics and Space Administration and the National Science Foundation.
\end{acknowledgements}
\bibliographystyle{aa}
\bibliography{harpsXXIX}

\end{document}